# *6AInets*: Harnessing artificial intelligence for the 6G network security: Impacts and Challenges

Navneet Kaur, *Department of Comp. Science, University of Missouri – St. Louis,* St. Louis, MO, USA, nk62v@umsystem.edu
Naresh Kshetri, *School of Business & Technology, Emporia State Uniersity.,* Emporia, KS, USA, nkshetri@emporia.edu
Purnendu Shekhar Pandey, *Department of Business Analytics*, *Jaipuria Institute of Management,* Lucknow, Uttar Pradesh, India, purnendu.pandey@jaipuria.ac.in

***Abstract*— This decade has witnessed the initiation of the digital revolution, as anticipated with the advent of 5G networks. Looking ahead to the 6G communication era, considerations are being made regarding how individuals will engage with the digital virtual world. The design of 6G technology, which will present enormous opportunities to develop and enhance human potential, will have a major impact on communications in the 2030s. We believe that in 6G we will see an unprecedented transformation that will set it apart from earlier wireless cellular network generations. Specifically, 6G will leverage ubiquitous AI services ranging from the network's core to its end devices, going beyond unpredictable limits. Despite the numerous advantages offered by 6G over existing technologies, there remains a pressing need to address security concerns. For example, the automation of critical processes in the 6G infrastructure will lead to a significantly broader and more intricate attack surface. Thus, the significance of Artificial Intelligence (AI) in providing security aspects within the envisioned 6G paradigm is substantial, but its integration presents a dual-edged dynamic. Therefore, to strengthen and validate the relevance of AI in securing 6G networks, this article elucidates how AI can be strategically used in 6G security, addressing potential challenges, and proposing solutions to enhance its role in securing networks.***

***Keywords—6G networks, 6G security, artificial intelligence, network security, solutions, wireless cellular***

## I. Introduction

New communication generations are revolutionizing connectivity, accessibility, and user experience, redefining how we engage with the digital world. These developments affect businesses, civilizations, and personal lives, redefining the digital landscape. 6G, a technology currently in the research and development stage, has the potential to transform how we engage with technology and the environment. The 2030s communication landscape will be significantly influenced by 6G technology, offering vast potential and opportunities to elevate and extend human capabilities [1]. However, what precisely is a 6G network? Furthermore, what distinguishes it from 5G networks?

6G, denoting the Sixth Generation of wireless communication technology, is poised to revolutionize the landscape by delivering unprecedented speed, reliability, and security [2]. Unlike its predecessor, 5G, 6G is not merely an evolutionary step but a paradigm shift, introducing transformative features beyond incremental improvements. It adopts a service-centric approach, advancing current technologies to create a more integrated digital world [3]. It promises ultra-low latency, higher data rates, enhanced spectral efficiency, faster speeds, more connected devices, and superior security features. 6G uses higher frequency bands, including terahertz frequencies, for faster data transmission rates and broader bandwidth [4]. Advanced antenna technologies like intelligent reflecting surfaces and MIMO systems enhance coverage, reliability, and energy efficiency [4]. Its applications extend beyond traditional communication to immersive experiences like holographic and haptic communication, augmented reality, and virtual reality, facilitating complex scenarios like remote surgery, autonomous vehicles, and smart city infrastructures [5]. This represents a significant advancement towards the future. In essence, 6G represents a significant advancement towards a future that is more intelligent, efficient, and connected than 5G was.

6G technology, which integrates Artificial Intelligence, will revolutionize communication by enabling smarter networks that adapt and optimize without human intervention [6]. Users can expect personalized experiences, smart living with AI-driven automation, and advanced industry automation. AI-empowered 6G networks will improve services like autonomous driving, robotics, and smart agriculture without human intervention [4]. However, these networks also present security challenges [4], including attacks on AI training procedures. An attacker could introduce fake data, render decision boundaries meaningless, and potentially infect other clients' models in federated learning [7]. Hence, while creating and deploying 6G technologies, security solutions should be tailored to network heterogeneity and dynamic conditions [8], balancing AI use for security and managing inherent hazards. Therefore, this study investigates the complex interactions between AI and 6G network security, highlighting the potential of AI to improve security and discussing challenges associated with integrating AI and ML in 6G networks.

The following sections of this paper are arranged as follows. 6G network vision and architectural requirements, along with the panorama of security threats are introduced in Section 2. Section 3 discusses the security implications of the next generation (6G) network. The current Literature Review is presented in Section 4. Section 5 delves into the contribution of AI towards strengthening the 6G network, while Section 6 examines some of the concerns and challenges associated with adopting AI in 6G. The paper's conclusion is provided in Section 7. Through in-depth conversations, we hope to paint a thorough picture of both current research projects and potential future research areas.

## II. Background

### A. Necessary Progression Beyond 5G Connectivity – Why 6G is needed?

Every generation of wireless technology has brought about revolutionary developments in the constantly changing field of telecommunications, revolutionizing the ways in which we connect and interact. The excitement about what lies ahead is already building as 5G networks continue to be deployed throughout the world. The next generation of wireless communication, or 6G, is about to change connection in ways that were previously thought to be unthinkable [6]. In spite of the existence of 5G, this section explains the strong arguments for why 6G is essential rather than merely a luxury.

The growth of the Internet of Things (IoT) and the emergence of industry 4.0 have led to a constant 1000x increase in network capacity. The emergence of new use cases and applications, such deep sea and space tourism, holographic telepresence, collaborative robotics, and the Internet of Everything (IoE), are pushing the limits of existing networks [9]. Undoubtedly, 5G was standardized and launched to support heterogeneous IoT applications through its high-frequency millimeter-wave (mmWave) functions, and rate-sensitive eMBB services, yet its ability to deliver smart city and smart utility services is still up for debate. Nevertheless, an unprecedented wave of cutting-edge Internet of Things products and services continues to emerge [5], such as haptics, drones, brain-computer interfaces, telemedicine, extended reality (XR) services, and networked autonomous systems leading to a shift from rate-centric enhanced mobile broadband (eMBB) services to ultra-reliable, low latency communications (URLLC) [6]. Moreover, other emerging technologies that haven't been utilized in telecoms up to this point—like smart wearables, implants, XR systems, haptics, flying cars [6], etc.—are anticipated to be employed in 6G. Furthermore, security technologies (like blockchain for data validation) and artificial intelligence (AI) algorithms (like network monitoring, data-driven business choices, preventative maintenance, fraud detection, etc.) are crucial to the realization of 6G [4].

Thus, the sixth generation (6G) cellular network, which is expected to be the next generation of mobile communications networks, is needed to improve thing-to-thing communication with the intelligent and self-sustaining orchestration systems [10]. In a nutshell, as our things get smarter and our industries become more advanced, 5G is working hard to keep up with the growing demand for better and faster connections. The challenge is not just about meeting the needs of today but also getting ready for all the exciting new technologies that will shape our future. It's like making sure our internet is ready for everything the future might bring.

### B. Related Work

The 6G networks, enabling technologies, designs, and open research challenges have been the subject of several studies in recent years. For example, In [11] the author explores the use of AI and ML in B5G network design and operation, examining recent advancements and future challenges in wireless network optimization. In [12] the authors highlight the potential of AI in enhancing security in 5G networks, but also warn of potential security risks if exploited by malicious actors. The authors in [13] reviews current research on AI in cellular networks, identifies obstacles, and presents a roadmap for achieving its potential. It also highlights future research directions, top challenges, and outlines a possible roadmap for Beyond 5G and 6G networks. [15] addresses 10 new issues from the viewpoints of computing, networking, and communication that will arise in integrating intelligent machine learning into the next generation 6G network.

[4] explores the opportunities and challenges of AI-based security and privacy provision in 6G systems, identifies future research directions, discusses challenges in AI-based security and privacy provision, and proposes viable solutions. Along with that [16] explores the security and privacy implications of 6G wireless systems, potential challenges with different technologies, and potential solutions. The paper [17] examines AI-based methods for developing 6G networks that strike a balance between security and service performance and offers guidance to researchers on how to build an ecosystem that combines both services and security. Researchers [18] discussed a few challenges for AI/ML and wireless communications, focusing on sixth-generation (6G) wireless networks. These challenges include computation in AI, distributed neural networks and learning, and semantic communications. [19] explores the security and privacy of AI-enabled-6G architecture, discussing threats like malicious cheating, low-quality local model training, and privacy stealing, and suggesting future research directions. The authors in [21] discuss the importance of AI-enabled NGNs in softwarized networks, highlighting their benefits, design requirements, and challenges. They present a use case study and experiments on an edge intelligence framework that orchestrates and deploys AI microservices using multisite cloud/edge-native NGNs.

These authors advocated several defense measures, to increase resilience to 6G network security threats, but none of them provide an all-in-one solution. Our purpose of the study is to highlight prospective pain areas and provide an overview of relevant obstacles along with AI mitigation strategies.

## C. Necessity of Security in the 6G Revolution

Security is a paramount consideration in the development and deployment of 6G (Sixth Generation) networks for several crucial reasons. Providing the top 10 reasons for security needs in 6G (as shown in Figure I below).

In summary, security stands as a foundational requirement for the successful deployment and adoption of 6G technology. It encompasses protection against a wide range of threats, including cyberattacks, data breaches, and unauthorized access, and is crucial for ensuring the integrity, privacy, and reliability of 6G networks and the diverse applications they support.

## III. ROLE OF AI IN PROVIDING SECURITY IN 6G HORIZON

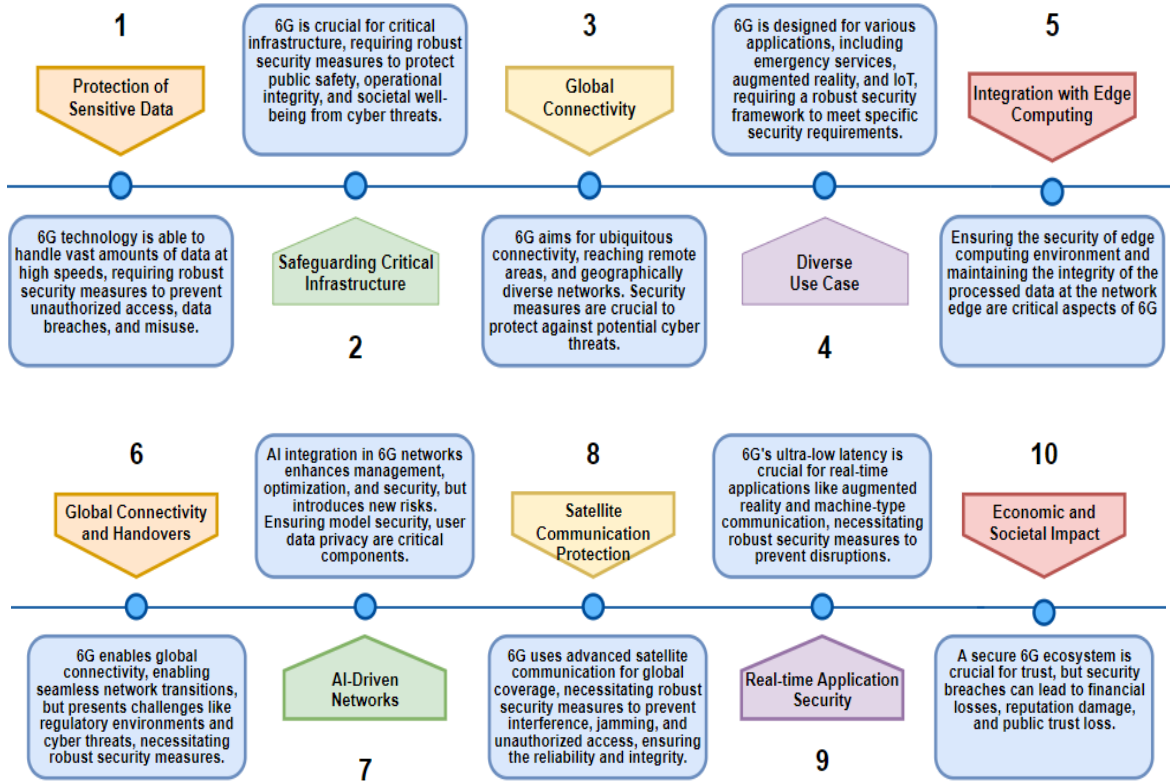

Fig. 1. Top 10 security requirements for 6G (Sixth Generation) network

The 6G revolution presents numerous challenges for innovation, but artificial intelligence (AI) is emerging as a revolutionary force to push the boundaries of what is possible. AI is preparing to explore the unexplored regions of the 6G horizon, overcoming the challenges of handling the size and complexity of anticipated networks. Deeper exploration of the 6G environment reveals that AI is the key to unlocking previously unimaginable possibilities [4]. Cyberattacks and real-time data integrity are becoming increasingly difficult, making AI-driven security solutions crucial for protecting networks from evolving threats and constantly monitoring for abnormalities and potential assaults [7]. AI's flexibility and security policies result in a more secure and coherent global network making unparalleled innovation and progress. Though we did have the same level of security concerns in 5G or the previous generation, at that time, AI was not as advanced & powerful as it is now. Thus, Artificial Intelligence (AI) is a key component and enabler of 6G networks, helping to transform communication. Below are some of the areas where AI techniques would be beneficial to make the next generation network stronger.

AI, with its adaptive and learning capabilities, can significantly contribute to overcoming security challenges in 6G, providing dynamic and proactive solutions across diverse architectural aspects.

## IV. CHALLENGES OF INTEGRATING AND ADOPTING AI FOR 6G NETWORKS

The partnership between AI-assisted future networks and AI can be a double-edged sword, as AI can potentially infringe on security, ethics, transparency, and privacy, and can also be used for strategic strikes. This section discusses these issues and suggests potential remedies.

## A. Cost and Efficiency

AI adoption often faces challenges due to costs like hardware, software, and operating expenses. The time and expertise needed to build AI algorithms and apps can increase development and implementation costs. Large-scale

6G networks may face efficiency issues due to model size, training time, and slow interpretation [19]. Organizations can reduce costs by using open-source software, edge-to-cloud cooperation, and their current network infrastructure.

As WWW transports towards bulk adoption of HTTPS encryption, cybercrime has pop up as big security loophole and threat to security industry, network industry, and the worldwide software industry. The countermeasures with blockchain technology can save cost and increase efficiency in 6G networks via data tampering prevention, malicious actor detection, Virtual Machine (VM) authentication, DoS attack prevention, security of IoT devices, multi-confirmation for attacks, solving security issues in Virtual Private Networks (VPNs), tracking internet delivery, integrity of data, resilient communication against critical systems, chaining open standard etc. [20].

### B. Complexity & Integration Challenges

Implementing complex AI models on edge devices can be challenging due to limited processing power and logistical challenges [2]. The integration challenges arising from differences in hardware, software, and communication protocols, as well as from the programming language and framework requirements, that work with specific functions or systems are some of the basic hurdles . To address these issues, model optimization approaches can be used. Another issue is Interoperability which is crucial in advanced cellular networks to prevent vendor lock-in and improve performance. Emerging technologies on digital landscape has significant AI security & privacy challenges and disparities between AI modules can increase signaling overhead and make it difficult to identify vendors responsible for decreased key performance indicators (KPI) [24] [26]. The open-source community has developed communication protocols, industry standards, and integration frameworks to enhance interoperability and standardization. Examples include Open Neural Network Exchange (ONNX) [27] for deep learning models and Synaptics' chips for wireless technologies.

TABLE I. Artificial Intelligence in mitigating the Security Risks Associated with 6G Landscape

| 6G Technologies | Description | Security Challenges | Proposed AI mitigation strategies |
|---|---|---|---|
| Terahertz Frequencies | 6G uses terahertz frequencies to boost data rates, but challenges include signal absorption and propagation issues. [9] | Terahertz frequencies pose security risks due to increased vulnerability to eavesdropping and interception. [16] | AI can improve security by dynamically adjusting encryption algorithms based on real-time threat assessments and can detect anomalies and potential attacks. |
| Massive MIMO and Beamforming | 6G builds on 5G technologies to enhance MIMO systems with advanced beamforming techniques, potentially reducing interference and improving spectral efficiency [2]. | Massive MIMO and Beamforming pose security risks due to potential signal interception and unauthorized access [21]. | AI-powered intrusion detection systems can swiftly identify and respond to security threats, making them more secure against unauthorized access. |
| Artificial Intelligence (AI) Integration | AI is expected to enhance network orchestration, optimization, and resource allocation for real-time decision-making, user behavior prediction, and dynamic network parameter adaptation [4]. | AI integration raises security concerns due to potential vulnerabilities in algorithms [4]. | AI enhances security by detecting and preventing attacks on models and using privacy-preserving techniques like federated learning to protect user data. |
| Network Slicing | 6G's network slicing technology is expected to enhance the creation of customized virtual networks for specific applications or industries, requiring fine-grained resource allocation and isolation [2]. | Cross-slice, and DoS attacks during authentication and authorization mechanisms are serious security concerns [22]. | AI can continuously monitor anomalies and potential breaches, guiding adaptive access control mechanisms and dynamically respond to changing security postures within individual slices. |
| Satellite Integration | Satellite communication capabilities, including low Earth orbit satellite | Satellite integration poses security risks like signal jamming | AI can analyze communication patterns to detect potential |

| | constellations, aim to enhance global coverage, minimize latency, and ensure reliable communication in remote areas [2]. | due to unauthorized access [23]. | interference or jamming attempts, and adjust encryption protocols based on detected threats. |
|---|---|---|---|
| Holographic Communication, Metaverse and Extended Reality (XR) | 6G aims to enable Metaverse, holographic communication, and immersive XR experiences with high data rates and low latency for seamless, interactive applications [2]. | These experiences raise security concerns, especially regarding data integrity and confidentiality. | AI can improve security by monitoring data integrity and detecting tampering attempts and implementing adaptive encryption for secure data transmission. |

*C. Data Governance and Management*

Data governance is crucial for AI-integrated edge devices, ensuring compliance with rules and guidelines [8]. Edge devices often collect noisy or imperfect real-time data, causing poor performance and inaccurate model predictions. Security and privacy concerns arise as data is migrated, impacting efficiency, storage, and processing needs. Researchers are utilizing Federated learning, Huffman coding, LZW compression, and blockchain-based frameworks like Ethereum and Hyperledger Fabric to improve data quality, security, and storage efficiency, while ensuring decentralized data management. Blockchain technology can affect the network security world including large and small businesses, banks, credit unions, retail giants that forms the solid foundation of data governance, P2P network, network verifiability as a ground-breaking technology. The use of blockchain to issue and validate certificates in networks, trusted room for retrieving information, features of decentralization, immutability, consensus mechanism for 6G environments [14].

*D. Security*

The integration of AI in 6G cloud environments presents significant security risks, including data breaches, cyberattacks, and privacy violations. Evasion attacks and poisoning assaults can cause misclassification and resource misprediction [16], while physical assaults can cause communication outages and hinder data processing. AI/ML technologies like adversarial training and defensive distillation improve resilience against attacks by introducing attack-like samples into training data. Therefore, security protocols must be considered from the beginning in AI system architecture, including code reviews and threat modeling. The security and safety of AI-based healthcare systems that are critical in nature can be improved via blockchain's distributed record keeping as a cryptographic hash and a decentralized network which makes digital assets unalterable and transparent [28].

*E. Latency*

Latency is a significant issue that can significantly impact AI systems in cloud environments. It can affect the usefulness and efficacy of the precision and promptness of AI predictions by creating a temporal lag between data collection and forecasting, affecting applications like autonomous driving and medical diagnostics [2]. AI systems employ federated learning, edge computing, and edge caching to address latency issues. Edge caching keeps data closer to the network's edge, while edge computing reduces processing latency. Decision trees and rule-based systems enhance real-time prediction responsiveness, while federated learning minimizes output delay and data privacy.

*F. Scalability & Power Consumption*

Scalability has an impact on the performance and adaptability of the system. Without sacrificing performance, the system need to be able to manage the expanding quantities of data and the increasing number of IoT devices. By maximizing load balancing, distributed computing, and parallel processing AI system performance and reliability can be enhanced. Another significant issue is high power consumption due to the need for advanced processing capabilities, high-performance CPUs, memory, and storage devices [29]. Energy-efficient technology, edge computing devices like ARM-based CPUs, optimizing AI algorithms, and reducing pointless data transfers, can reduce energy consumption. Combining these technologies with renewable energy sources [6] can also reduce carbon impact.

*G. Transparency*

Though there has been a lot of work already done in creating robust AI models in the space of IoT edge enabled network that not only helps in delivering a better infrastructure but also addresses key concerns with privacy & security but still lacks the feature that explain how those models reached to the conclusion based on which they took that action. Explainable AI models [6] could help bridge the gap. Apart from AI models for IoT network, there has been the use of blockchain technology in managing the transparency and trust for public service delivery like the 6G networks to uplift the living of standards of people and improve security [30].

## V. Conclusion and Future Work

The 21st century is poised for a wireless communications renaissance, propelled by AI. This article has meticulously reviewed recent research areas, spotlighted significant challenges, and outlined a roadmap to realize AI's potential in cellular networks. We've provided an overview of current AI approaches and identified their limitations. Moving forward, we plan to leverage and evaluate AI model performance using Explainable AI (XAI) and conduct extensive experiments to assess XAI's potential in enhancing the credibility of AI-assisted methodologies for future networks. Overcoming technological obstacles in this domain should drive fundamental research and engineering creativity, paving the way for realizing the AI vision in Beyond-5G and 6G cellular networks.

In conclusion, the security landscape of 6G networks is intricate, marked by challenges stemming from complex architecture and diverse technologies. To tackle these challenges, a comprehensive security framework is indispensable, with AI emerging as a linchpin in fortifying 6G networks against known and unforeseen security threats. As 6G continues its advancement, a holistic approach integrating advanced encryption, authentication mechanisms, and AI-driven security measures will be pivotal for ensuring the integrity and resilience of these next-generation wireless networks and to guarantee the integrity and confidentiality of data in AI-enabled 6G networks, future research should examine privacy-preserving machine learning algorithms, safe federated learning methodologies, and strong cybersecurity measures.